# Crystal structure and properties of iron-based spin-chain compound $Ba_9Fe_3Se_{15}$

Jun Zhang[†1], Alexander C. Komarek[†2], Meiling Jin[†3], Xiancheng Wang[†*1,4], Yating Jia[1], Jianfa Zhao[1], Wenmin Li[1], Zhiwei Hu[2], Wei Peng[2], L.H. Tjeng[2], Zheng Deng[1], Runze Yu[1], Shaomin Feng[1], Sijia Zhang[1], Min Liu[1,4], Yi-feng Yang[1,4,5], Hong-ji Lin[6], Chien-Te Chen[6], Xiaodong Li[7], Jinlong Zhu*[3] and Changqing Jin*[1,4,5]

[1] Beijing National Laboratory for Condensed Matter Physics, Institute of Physics, Chinese Academy of Sciences, Beijing 100190, China.

[2] Max Plank Institute for Chemical Physics of Solids, Nöthnitzer Str. 40, D-01187 Dresden, Germany

[3] Department of Physics, Southern University of Science and Technology, Shenzhen 518055, China

[4] School of Physics, University of Chinese Academy of Sciences, Beijing 100190, China

[5] Materials Research Lab at Songshan Lake, Dongguan 523808, China.

[6] National Synchrotron Radiation Research Center (NSRRC), 101 Hsin-Ann Road, Hsinchu 30076, Taiwan.

[7] Institute of High Energy Physics, Chinese Academy of Sciences, Beijing 100049, China

*Corresponding author: wangxiancheng@iphy.ac.cn; zhujl@sustech.edu.cn; jin@iphy.ac.cn

† These authors contributed equally

**Keywords:** iron-based compound; spin chain; high pressure; spin state

**Abstract:** We report the synthesis of a new quasi one-dimensional (1D) iron selenide. $Ba_9Fe_3Se_{15}$ was synthesized at high temperature and high pressure of 5.5 GPa and systematically studied via structural, magnetic and transport measurements at ambient and at high-pressures. $Ba_9Fe_3Se_{15}$ crystallizes in a monoclinic structure and consists of face-sharing $FeSe_6$ octahedral chains along the *c* axis. At ambient pressure it exhibits an insulating behavior with a band gap ~460 meV and undergoes a ferrimagnet-like phase transition at 14 K. Under high pressure, a complete metallization occurs at ~29 GPa, which is accompanied by a spin state crossover from high spin (HS) state to low spin (LS) state. The LS appears for pressures P >36 GPa.

# Introduction

The iron selenide compounds have attracted numerous attentions because of the superconductivity (SC) reported in the iron-based superconductors (IBSs) [1-5]. FeSe is a simple layered superconductor where the antifluorite-type [FeSe] layers are bonded with Van der Waals force. The bulk FeSe exhibits SC below ~9 K, while the SC transition temperature $T_c$ can be enhanced by introducing charge carriers into the system either via the interface effect between the mono-layer of FeSe and the $SrTiO_3$ substrate or intercalating alkaline-earth metals into the FeSe layers. Although FeSe does not possesses long-range magnetic ordering, the stripe spin fluctuation was evidenced to interplay with the SC in FeSe, which reveals that the SC is driven by the spin fluctuation. Besides the layered iron selenide superconductors, it's inspired that the SCs were observed in the spin-ladder iron chalcogenides $BaFe_2S_3$ and $BaFe_2Se_3$ under high pressure [6-8]. The crystal structure of these iron chalcogenides consists of spin ladders along the $c$ axis formed by the edge-sharing $FeS_4$ ($FeSe_4$) tetrahedrons. $BaFe_2S_3$ is a Mott insulator with stripe-type antiferromagnetic ordering below 120 K. By applying pressure the superconductivity with $T_c$ ~ 24 K was induced above the critical pressure of 10 GPa, before which the antiferromagnetic ordering was suppressed [6]. While $BaFe_2Se_3$ undergoes a block-type antiferromagnetic structure at $T_N$ = 256 K [9]. The SC ($T_c$ ~ 11 K) emerges after the local magnetic moment is decreased by pressure and comparable to the layered IBSs [8]. It seems that the SC discovered in IBSs is closely related with the magnetism although they have various antiferromagnetic structures.

It will be interesting to look for other new IBSs with quasi one-dimensional (1D) structure. Using high-pressure method we have synthesized several new hexagonal $Hf_5Sn_3Cu$-anti type materials with a general formula of $A_3BX_5$, where the face-sharing $BX_6$ octahedral chains are separated by a large distance more than 9 Å, exhibiting a strong 1D structure characteristic [10-15]. Among these quasi 1D materials, the iron telluride $Ba_9Fe_3Te_{15}$ with $FeTe_6$ octahedral chains was reported to display a 1D antiferromagnetic behavior [15] and show pressure-induced SC with a maximum $T_c$ ~4.7 K [16]. Here, we reported the synthesis of new iron selenide $Ba_9Fe_3Se_{15}$. The crystal structure was solved by a combined powder and single crystal X-ray diffraction (XRD) method. $Ba_9Fe_3Se_{15}$ crystallizes in a monoclinic structure with face-sharing $FeSe_6$ octahedral chains along the $c$ axis. The magnetic susceptibility measurements suggest a ferrimagnet-like phase transition at ~14 K. $Ba_9Fe_3Se_{15}$ is an insulator at ambient pressure with a band gap ~460 meV. When applying high pressure, a complete pressure-induced metallization and a spin state transition (SST) occur simultaneously at ~29 GPa. The SST is finished when P>36 GPa and no SC was observed within the experimental pressure range of 52 GPa.

**Experiments**

$Ba_9Fe_3Se_{15}$ was synthesized by solid state reaction at high pressure and high temperature conditions. Lumps of Ba (Alfa, immersed in oil, >99.2% pure), crystalline powders of Fe (Alfa, 99.998% pure) and Se (Alfa, >99.999% pure) were used as the starting materials. Initially, a precursor of BaSe was prepared by heating the mixture of Ba blocks and Se powder in an evacuated quartz tube at 600 °C for 10 hours. A mixture of BaSe, Fe and Se powders with stoichiometric ratio of 3:1:2 was ground and pressed into a flake, which was then sintered at 1200 °C under 5.5 GPa for 30 minutes. A black polycrystalline sample of $Ba_9Fe_3Se_{15}$ was obtained. Small single crystals could be extracted from the polycrystalline sample.

Single crystal XRD measurements have been performed on a *Bruker D8 VENTURE* single crystal X-ray diffractometer using monochromatic Mo *Kα* radiation (λ = 0.71073 Å) and a bent graphite monochromator for about 3×intensity enhancement as well as a *Photon* CMOS large area detector. 91899 reflections [with 99.96% (98.09%) coverage up to sinθ/λ= 0.9052 (0.9552) ] have been collected up to $2θ_{max}$= 85.52 ° with a redundancy of 4.374 and an internal R-value of 3.96%. Powder XRD measurements have been performed on a high resolution *Bruker D8 Discover A25* powder X-ray diffractometer with monochromatic Cu-$K_{α,1}$ radiation.

The stoichiometry of single crystal $Ba_9Fe_3Se_{15}$ was determined by energy dispersive X-ray spectroscopy (EDX). A physical property measurement system (PPMS) was used for measuring the resistivity under ambient pressure by a standard four-probe technique. The Mössbauer spectroscopy measurements were performed using a SEE Co. conventional constant acceleration type spectrometer in transmission geometry with an $^{57}$Co(Rh) source at room temperature. To optimize the signal intensity, a $Ba_9Fe_3Se_{15}$ sample synthesized with $^{57}$Fe was used for the Mössbauer spectroscopy measurements. Soft X-ray absorption scattering (XAS) of the Fe-$L_{2,3}$ edges was measured at the beamline BL11A of the National Synchrotron Radiation Research Center (NSRRC) in Taiwan.

High-pressure synchrotron X-ray diffraction experiments were performed at the Beijing Synchrotron Radiation Facility at room-temperature with a wave-length 0.6199 Å at room-temperature. The high-pressure X-ray emission spectra (XES) experiments were performed at the 16 ID-D station of HPCAT of the Argonne National Laboratory. The in-situ resistance measurements were conducted in a Maglab system via four-probe electrical conductivity methods in DACs made of CuBe alloy. The single crystal sample was crushed into small fragments. A plate like piece with the size of 40×40×5 μm was pick up for the high-pressure resistance measurements. The details of in situ high-pressure experiments can been seen in the Ref.[16].

**Results**

The synthesized $Ba_9Fe_3Se_{15}$ single crystals are rod-like in shape with the typical size of ~0.3 mm in length, see the inset of Fig. 1(a). The stoichiometry of single crystalline $Ba_9Fe_3Se_{15}$ measured by EDX shows that the average atomic ratio of Ba: Fe: Se is 3.1(6):1:4.9(8) which is close to the stoichiometric formula of $Ba_9Fe_3Se_{15}$, as seen in Fig. 1(a).

$Ba_9Fe_3Se_{15}$ crystallizes in a monoclinic crystal structure that was solved within a combined powder and single crystal X-ray diffraction (XRD) study and that is related to the already known compound $Ba_9Fe_3S_{15}$ [17]. Powder X-ray diffraction measurements clearly indicate that neither hexagonal nor crystal structures with symmetries higher than monoclinic are able to describe the X-ray diffraction (XRD) patterns of $Ba_9Fe_3Se_{15}$ satisfactorily, see Fig. 1(b-e). The space group with highest symmetry that provides a structural model that is able to describe the powder and single crystal XRD data satisfactory is space group *C*2/*c*. A Rietveld refinement of the powder XRD data that is using the positional parameters obtained by single crystal XRD is shown in Fig. 1(f) and gives the lattice constant *a*=16.5947 Å, *b*=9.6128 Å and *c*=18.6735 Å. The summary of the crystallographic data is shown in Table I. The finally obtained crystal structure based on space group *C*2/*c* consists of face-sharing $FeSe_6$ octahedral chains running along the *c* axis that are separated by $Ba^{2+}$ ions and Se-chains with a distance larger than 9.5 Å, thus exhibiting strong 1D spin chain structure as seen in Fig. 1(g). There are two Fe sites in the monoclinic crystal structure of $Ba_9Fe_3Se_{15}$ (see Fig. 1(h)). The distance of nearest neighbour Fe1 and Fe2 is 3.1825 Å, while it is 2.9717 Å for Fe2-Fe2. The bond valence sums (BVS) of both Fe sites, Fe1 and Fe2, amounts to 2.02(1) which indicates a 2+ oxidation state of the iron ions in $Ba_9Fe_3Se_{15}$. For the Se ions the BVS formalism clearly indicates the lowest Se oxidation states for Se ions within the Se-chains. This indicates the presence of $Se^{1-}$ ions within these Se-chains whereas the valences of all the other Se ions within the $FeSe_6$ octahedral chains are distinctly larger (by ~0.7 in average) and, thus, indicative of $Se^{2-}$ ions. Finally, the $Fe^{2+}$ oxidation state in $Ba_9Fe_3Se_{15}$ is in agreement with the appearance of $Se^{1-}$ ions within the Se-chains since the 6 $Se^{1-}$ and 9 $Se^{2+}$ ions are in charge balance with the 3 $Fe^{2+}$ and 9 $Ba^{2+}$ ions.

The oxidation state of the iron ions in $Ba_9Fe_3Se_{15}$ has been investigated by soft X-ray absorption spectrum at the Fe-$L_{2,3}$ edge, which is very sensitive to the valence state [18, 19] and the local environment [20, 21]. Fig. 2(a) shows the Fe $L_{2,3}$ XAS of $Ba_9Fe_3Se_{15}$ together with those of $Mg_{0.96}Fe_{0.04}O$ as a $Fe^{2+}$ reference (taken from Ref.[22]) and $Fe_2O_3$ as a $Fe^{3+}$ reference. The spectrum of $Ba_9Fe_3Se_{15}$ is shifted by more than 1 eV to lower photon energy relative to that of the $Fe_2O_3$, but is at the same photon energy as that of $Mg_{0.96}Fe_{0.04}O$, demonstrating the 2+ valence state for Fe ions in $Ba_9Fe_3Se_{15}$. The oxidation state of iron in $Ba_9Fe_3Se_{15}$ was also confirmed by Mössbauer spectroscopy, see Fig. 2(b). Because there are two Fe sites in $Ba_9Fe_3Se_{15}$, two curves were used to fit for the measured data. The obtained isomer shift (CS) is about 0.91 mm/s and 0.92 mm/s for the Fe1 and Fe2 atoms respectively, which confirms the 2+ oxidation state observed in the BVS of both Fe atoms as derived from single crystal X-ray diffraction. The quadrupole splitting (QP) is 0.78 mm/s for Fe1 and 1.35 mm/s for Fe2. The magnitude of the QP reflects the homogenization of the field around the Fe atoms. The smaller value of the QP of the Fe1 atom indicates that its field is more homogenous than that of the Fe2 atom, which is consistent with the observation of dimerization of two neighboring Fe2 atoms. The dimerization should result in highly asymmetric selenium environments due to the shift of both Fe2 atoms towards the shared face of the two adjacent Se octahedral. In contrast to that the Fe1 atoms are located almost in the center of their Se octahedral surrounding.

Fig. 3(a) shows the temperature dependence of resistivity for $Ba_9Fe_3Se_{15}$ sample, which exhibits a semiconducting behavior. The inset presents the linear fit to the curve of ln(rho) as a function of inverse temperature using the formula of $\rho \propto \exp(\Delta_g/2k_BT)$, where $\Delta_g$ is the semiconducting band gap and $k_B$ is the Boltzmann's constant. The resistivity curve can be well fitted and the band gap $\Delta_g$ is evaluated to be ~460 meV. The resistivity of $Ba_9Fe_3Te_{15}$ is also presented in Fig. 3(a). Compared with its sister compounds $Ba_9Fe_3Te_{15}$, the band gap of Ba$_9$Fe$_3$Se$_{15}$ is enhanced when the anions of Te are replaced by smaller ion size of Se. This is a typical phenomenon observed in the system with quasi 1D conducting chains, where the electron hopping among the chains determines the transport property [23, 24]. Since the Se-4$p$ electrons are more localized than Te-5$p$ ones, the replacement of Te by smaller size of Se should reduce the electron hopping between FeX$_6$ chains and result in an increase of the band gap.

Fig. 3(b) displays the magnetic susceptibility measured for the polycrystalline sample of Ba$_9$Fe$_3$Se$_{15}$ with magnetic field of $H$=100 Oe. It demonstrates a ferrimagnet-like phase transition at ~14 K. The lower inset of Fig. 3(b) is the magnetic hysteresis loop measured at 2 K. The magnetization shows saturation when $H$ exceeds 500 Oe. The saturated moment is about 0.7 $\mu_B$/Fe which is much smaller than the expected for Fe$^{2+}$ with S=2. It suggests that the magnetic phase is ferrimagnetic or canted antiferromagnetic. The susceptibility in the paramagnetic region deviates from Curie-Weiss behavior within 300 K, as shown in the upper inset of Fig. 3(b). Therefore, we conducted a measurement within a higher temperature range of 500-700K and made a Curie-Weiss fit. The effective moment according to this fit amounts to $\mu_{eff}$ ~ 6.4 $\mu_B$ per Fe. This value is much larger than the expected value for Fe$^{2+}$ with a high spin state (HS, S=2), which indicates that the effective moment should contain orbital contributions.

We now resort to the element selective XMCD spectrum to further study the magnetic property of the Fe ions. An important feature of XMCD experiments is that there are sum rules, which were developed by Thole *et al.* [25] and Carra *et al.* [26], to determine the ratio between the orbital m$_{orb}$=$L_z$ and spin $m_{spin}$ = $2S_z$ contributions to the magnetic moment, namely,

$$\frac{Lz}{2Sz+7Tz} = \frac{2}{3}\frac{\int_{L_{2,3}}(\sigma^+ - \sigma^-)dE}{\int_{L_3}(\sigma^+ - \sigma^-)dE - 2\int_{L_2}(\sigma^+ - \sigma^-)dE} \quad (1)$$

where the $\sigma^+$ and $\sigma^-$ indicate the spectra taken with circularly polarized X-rays with the photon spin parallel or antiparallel to the applied magnetic field, respectively. $T_z$ denotes the magnetic dipole moment. For ions in octahedral symmetry $T_z$ is small and can be neglected compared to $S_z$ [27]. Figure 3(c) shows the Fe $L_{2,3}$ XMCD spectrum of Ba$_9$Fe$_3$Se$_{15}$ taken at 25 K under 6 T. The spectra were taken with the photon spin parallel $\sigma^+$ (red) and antiparallel $\sigma^-$ (black) to the magnetic field. One can clearly observe large differences between the two spectra. Their difference is the XMCD spectrum (blue).

In this particular case, we can immediately recognize the presence of a large orbital moment in Fig. 3(c) since there is a large net negative integrated XMCD spectral weight. By using Eq. 1, we find $m_{orb}/m_{spin} = 0.35$ and thereby the orbital moment was calculated to be ~1.4 $\mu_B$ if assuming S = 2 for HS $Fe^{2+}$. Using the orbital and spin momentum values, we can calculate the g-factor to be ~ 1.58 and then the effective moment $\mu_{eff}$ ~ 6.1 $\mu_B$ via the equation of $\mu_{eff} = g\sqrt{J(J+1)}$, where J = 3.4 $\mu_B$ is the sum of L and S. The obtained $\mu_{eff}$ from the XMCD experiments agrees well with that from magnetic susceptibility measurement, which confirms the unquenched orbital moment in the system and the assumption of the HS form with S=2. The observation of sizeable orbital moment is quite common in cobalt based compounds with a high spin $3d^6$ ionic configuration in octahedral coordination [28-30]. Also several iron based compounds such as $FeBr_2$ [31, 32] and $Ba_9Fe_3Te_{15}$ [15] have been reported to possess unquenched orbital moment.

The properties of $Ba_9Fe_3Se_{15}$ under high pressure condition was investigated. The *in-situ* angle dispersive synchrotron XRD patterns of $Ba_9Fe_3Se_{15}$ were recorded to study the structure stability under pressure, as shown in Fig. 4(a-c). Fig. 4(a) is the XRD patterns measured under pressure within 51 GPa. All the Brag peaks shift gradually to high angle direction as the increase of pressure, suggesting the shrink of lattice constants under pressure. No new Brag peak is observed, which reveals that the crystal structure is stable within the highest experimental pressure. Fig. 4(b) presents the pressure dependence of the lattice parameters *a*, *b* and *c*. These lattice parameters are decreased by ~13.2%, ~14.0% and ~17.4% within 51 GPa, respectively. In addition, an anomaly can be clearly seen at ~30 GPa for each curve. Since the crystal structure is stable, the anomaly suggests an electric transition, which will be confirmed by the following experiments. The cell volume as a function of pressure is plotted in Fig. 4(c). Using the Birch–Murnaghan equation, the data of P<30 GPa are fitted; the bulk modules $B_0$=35 GPa is obtained.

For the transition-metal atom, the Kβ (3p→1s) emission line is extremely sensitive to its spin state. Fig. 5(a) displays the Fe Kβ XESs of $Ba_9Fe_3Se_{15}$ between ambient pressure and 36 GPa. All the XESs show a main peak called Fe $K\beta_{1,3}$ spectral line and a satellite peak denoted as Fe Kβ' line on the lower energy side of Fe $K\beta_{1,3}$. When pressure increases, the intensity of Fe $K\beta'$ peak becomes weak and almost disappears at 36 GPa. At the same time, its spectral weight is transferred to the main peak of Fe $K\beta_{1,3}$, as can be clearly seen in Fig. 5(b). The very weak Fe $K\beta'$ peak indicates a low spin state (LS, S=0) should be reached. The change of spin moment can also be estimated from the integrated absolute difference (IAD) between the Fe Kβ XESs. According to the procedure described by Vankó et [33], the spectra were normalized firstly with respect to the areas and then shifted to centers of mass at the same position. To obtain the IAD, we can use the spectrum taken at the highest experimental pressure of 36 GPa as the reference. Fig. 5(b) presents the typical normalized XES of Fe Kβ spectral line taken at 1 GPa and the difference of the spectral intensity. The integrals of the absolute difference under different pressure is presented in Fig. 5(c). At ambient pressure the value of IAD is obtained to be 0.12, which agrees with that expected for a complete HS to LS transition in the system containing $Co^{3+}$ ions [34, 35]. Since $Fe^{2+}$ has the same 3$d$ electrons with $Co^{3+}$, the 0.12 value means a HS (H=2) for the ambient pressure and LS (S=0) for the case of pressure exceeding 36 GPa. The IAD value decreases with pressure and drops quickly at 29 GPa, which could be caused by either the enhancement of the crystal field or the delocalization of 3$d$ electrons. The later will imply that the metallization plays a predominant role. The pressure dependence of IAD indicates that a spin state transition starts at about 29 GPa, which is consistent with the abnormal variation of the lattice parameter under pressure observed in the high-pressure XRD experiments.

The high-pressure electronic transport properties of $Ba_9Fe_3Se_{15}$ were investigated as shown in Fig. 6(a, b). The insulating characteristic is gradually suppressed by pressure and a pressure-induced metallization occurs stating from ~21 GPa, where it exhibits metallic behavior at low temperature and undergoes a metal to insulator transition (MIT) at ~130 K. As pressure increases to 26 GPa, the MIT temperature is enhanced to ~250 K. When pressure exceeds 29 GPa, $Ba_9Fe_3Se_{15}$ exhibits metallic behavior within measured temperature range. The complete metallization at ~29 GPa causes the delocalization of the 3$d$ electrons at room temperature, which agrees with the sharp decrease of the IAD value associated with localized spin moments. This MIT is usually observed in a quasi 1D conducting system, in which a high-dimensional metallic state will be realized after the interchain electric hopping $t$ exceeds a critical value of $t^*$. However, the interchain hopping can be masked by the thermal fluctuation as temperature increases, which should lead to a MIT. It means that the MIT temperature will increase as the interchain hopping is enhanced. For $Ba_9Fe_3Se_{15}$, the increase of MIT temperature with pressure implies the resistance we measured should have the contribution of the interchain hopping and demonstrates the quasi 1D conducting properties in $Ba_9Fe_3Se_{15}$. Within the experimental pressure range of 0-51 GPa, no superconductivity is observed.

**Conclusion**

In conclusion, we synthesized the new iron based compound $Ba_9Fe_3Se_{15}$ under high-pressure and high-temperature conditions. The crystal structure was solved by combining powder and single crystal XRD measurements. $Ba_9Fe_3Se_{15}$ crystallizes in a monoclinic *C*2/*c* structure, where the face-sharing $FeSe_6$ octahedral chains are separated by a distance larger than 9.5 Å. $Ba_9Fe_3Se_{15}$ possesses a ferrimagnet-like phase transition at ~14 K and has a sizable unquenched orbital moment. It is an insulator with a band gap ~460 meV at ambient pressure. When applying pressure, the metallization occurs starting at ~21 GPa and a complete metallization is realized as pressure exceeds 29 GPa. The pressure-induced spin state transition is related with the delocalization of 3*d* electrons and a low spin state is reached when pressure exceeds 36 GPa.

Acknowledgments

We greatly appreciate the support of the National Key R&D Program of China and the Natural Science Foundation of China under grant no. 2018YFA0305700, 11974410, 12004161, 11974397, 11921004 and 2017YFA0302900. We acknowledge the support from the Max Planck POSTECH Hsinchu Center for Complex Phase Materials.

Table 1 Atomic Coordinates and Isotropic Displacement Parameters ($Å^2$) of $Ba_9Fe_3Se_{15}$.

Empirical formula: $Ba_9Fe_3Se_{15}$
Space group: *C2/c* (15) – monoclinic
a=16.5947(2) Å, b= 9.6128(1) Å, c=18.6735(2) Å
α=90°, β= 90.110(1) °, γ=90°
V=2978.82(6) $Å^3$, Z=2
Goodness-of-fit: 1.42; $R/R_w$ values: 4.29%/6.83%

| Atom | *x* | *y* | *z* | U(eq) |
|---|---|---|---|---|
| Ba1 | 0.68375(9) | 0.29790(19) | 0.08279(8) | 0.0085(3) |
| Ba2 | 0.30719(9) | 0.3252(2) | 0.08345(8) | 0.0114(4) |
| Ba3 | 0.00884(12) | 0.37534(18) | 0.08328(9) | 0.0117(4) |
| Ba4 | 0.18913(10) | 0.1892(2) | 0.24974(11) | 0.0149(5) |
| Ba5 | 0.5 | 0.1229(3) | 0.25 | 0.0122(6) |
| Fe1 | 0 | 0 | 0 | 0.0197(19) |
| Fe2 | 0.0001(3) | 0.0007(6) | 0.17043(8) | 0.0205(10) |
| Se1 | 0.3334(2) | 0.0021(4) | 0.01606(5) | 0.0128(4) |
| Se2 | 0.88867(18) | 0.1206(3) | 0.08253(13) | 0.0094(7) |
| Se3 | 0.6669(2) | 0.0023(4) | 0.17336(5) | 0.0134(5) |
| Se4 | 0.11547(18) | 0.1084(3) | 0.08324(14) | 0.0112(7) |
| Se5 | 0.38701(16) | 0.3857(3) | 0.24968(16) | 0.0100(7) |
| Se6 | 0 | 0.2288(5) | 0.25 | 0.0156(11) |
| Se7 | 0.4960 (2) | 0.2704 (3) | 0.0834 (1) | 0.0111(7) |
| Se8 | 0.33280(4) | -0.0007 (3) | 0.15002(4) | 0.0128(4) |

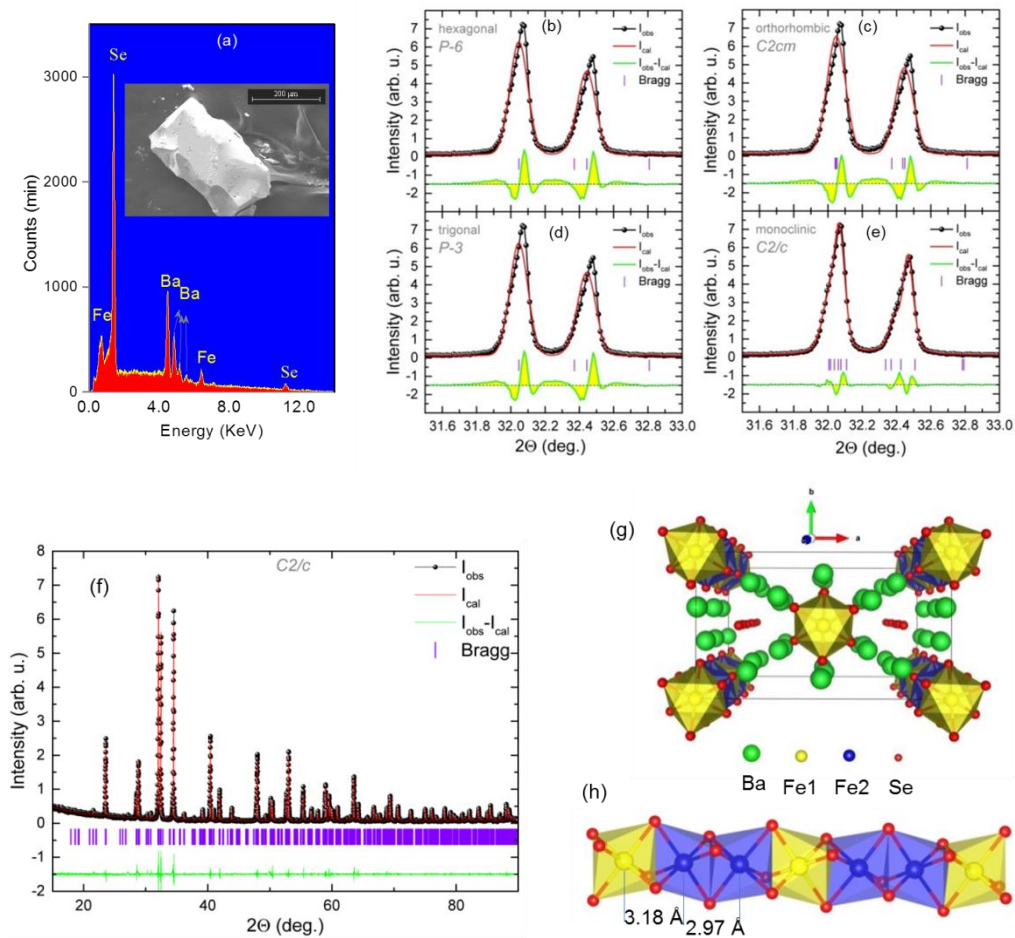

Fig. 1. (a) EDX spectrum measured for a $Ba_9Fe_3Se_{15}$ single crystal. The inset shows the Scanning Electronic Microscopy of $Ba_9Fe_3Se_{15}$ single crystal. (b-e) Rietveld fits of high resolution powder XRD patterns of $Ba_9Fe_3Se_{15}$ for different symmetries based on positional parameters obtained by crystal structure solutions based on single crystal XRD data. Here, a close-up of the two strongest structural reflections is shown. The green lines indicate the difference between observed and calculated intensities (shifted by -1.5 in $y$-direction). (f) The powder XRD and its refinement using $C2/c$ structure model. (g) The sketch of the crystal structure of $Ba_9Fe_3Se_{15}$ for the top view. (h) The face-sharing FeSe6 octahedral chain.

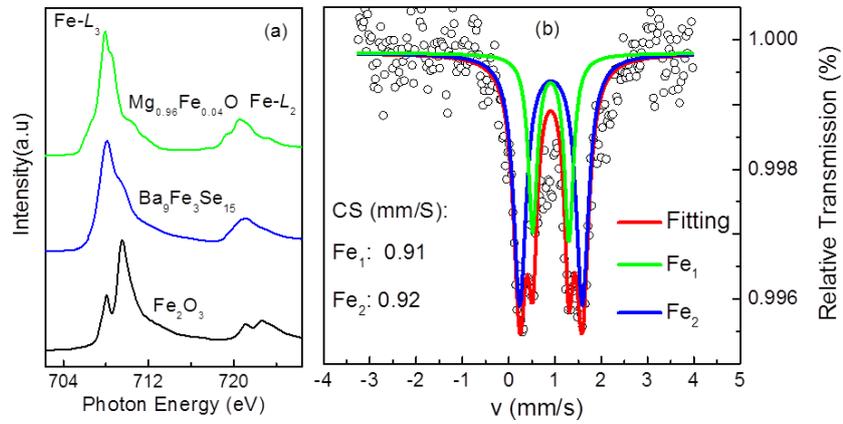

Fig. 2 (a) Fe $L_{2,3}$ XAS spectra of $Ba_9Fe_3Se_{15}$ and of $Fe_2O_3$ and $Mg_{0.96}Fe_{0.04}O$ (from Ref.[22]) as a $Fe^{3+}$ and a $Fe^{2+}$ references, respectively; (b) Mössbauer spectroscopy for $Ba_9Fe_3Se_{15}$ with isotope of $^{57}Fe$.

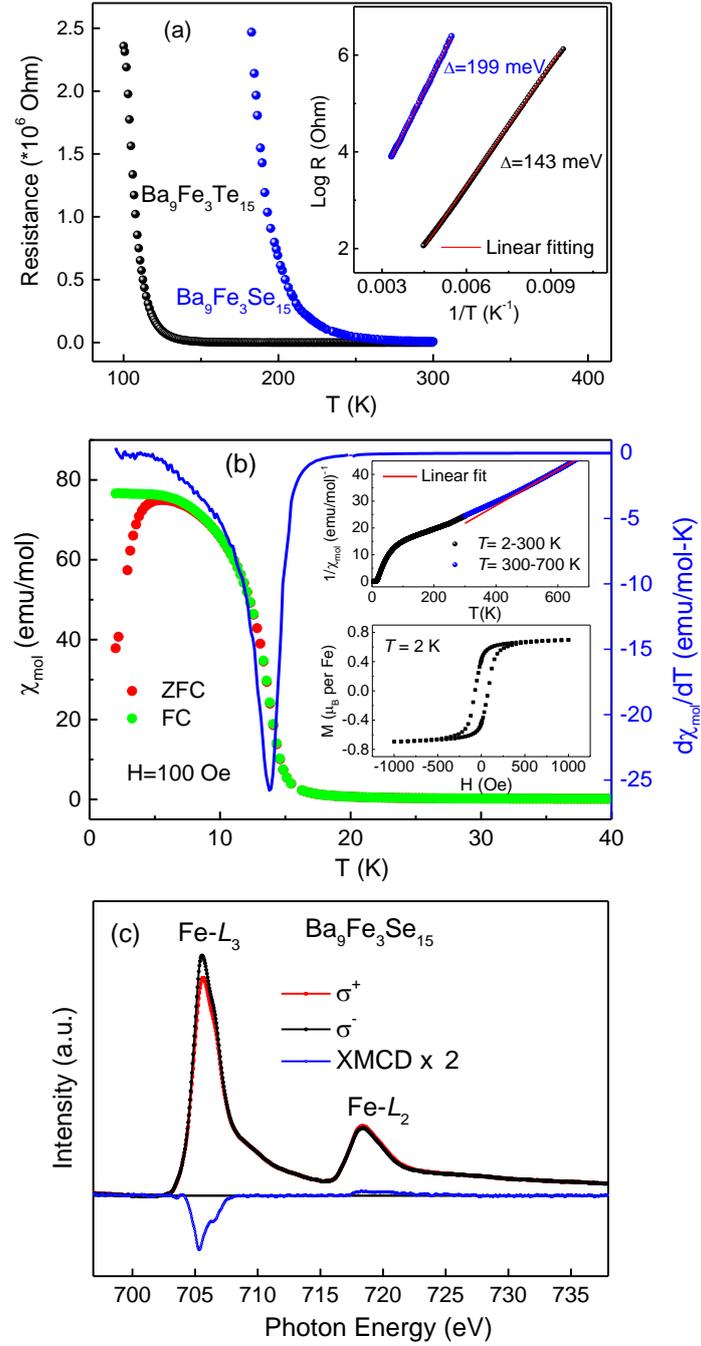

Fig. 3 (a) Temperature dependence resistivity of $Ba_9Fe_3X_{15}$, X= Se and Te. The inset is the ln(rho) as a function of inverse temperature and its linear fitting. (b) The susceptibility measured with H=100 Oe and the temperature derivative susceptibility showing the magnetic transition temperature $T_N$~14 K. The upper inset is the inverse susceptibility within the temperature range of 2-700 K and the lower inset is the magnetic hysteresis loops measured at 2 K. (c) Fe-$L_{2,3}$ XMCD spectra of $Ba_9Fe_3Se_{15}$ with the photon spin parallel ($\sigma^+$ red line) and antiparallel ($\sigma^-$ black line) to the applied magnetic field, respectively. The difference spectrum (XMCD) is shown in blue line.

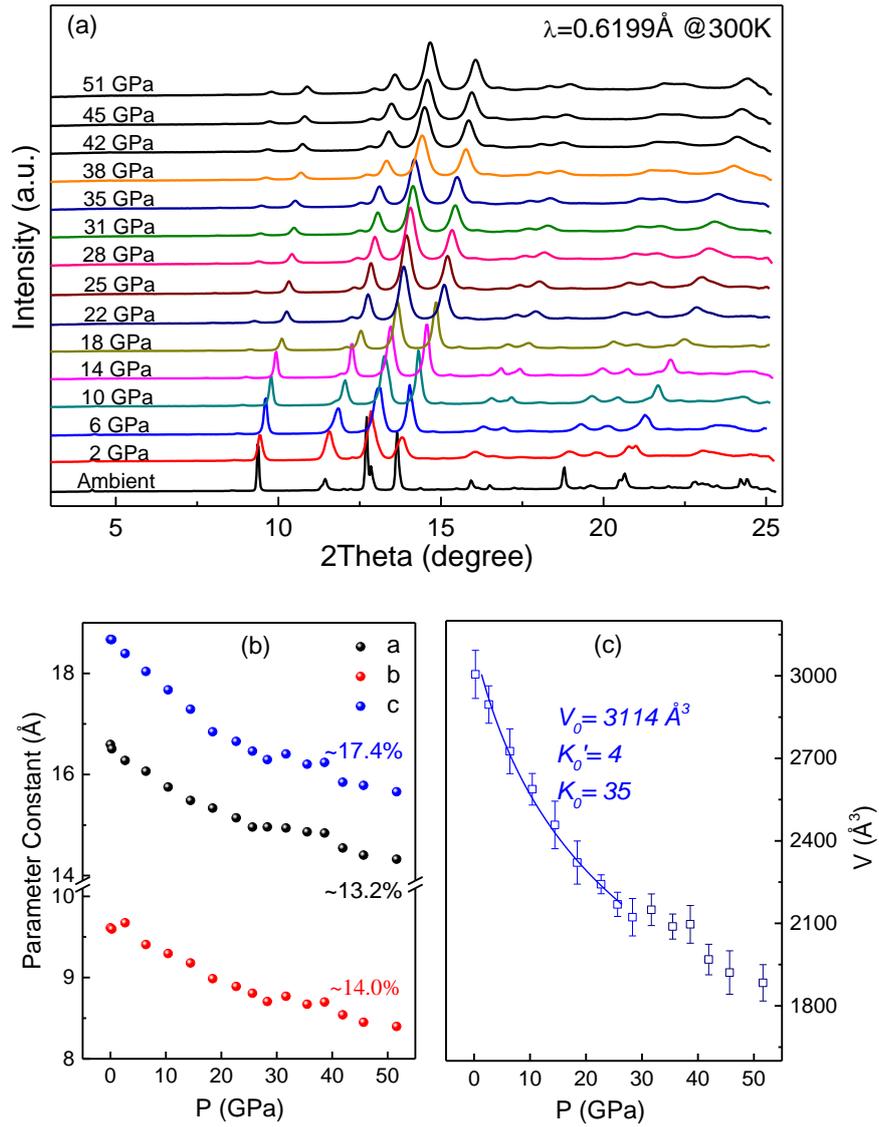

Fig. 4 (a) The synchrotron X-ray diffraction patterns of $Ba_9Fe_3Se_{15}$ measured at room temperature and high pressure conditions. (b) The lattice parameter *a*, *b*, *c* as a function of pressure. (c) The pressure dependence of unit volume *V*. The solid line is the fitting using the Birch–Murnaghan equation within 30 GPa.

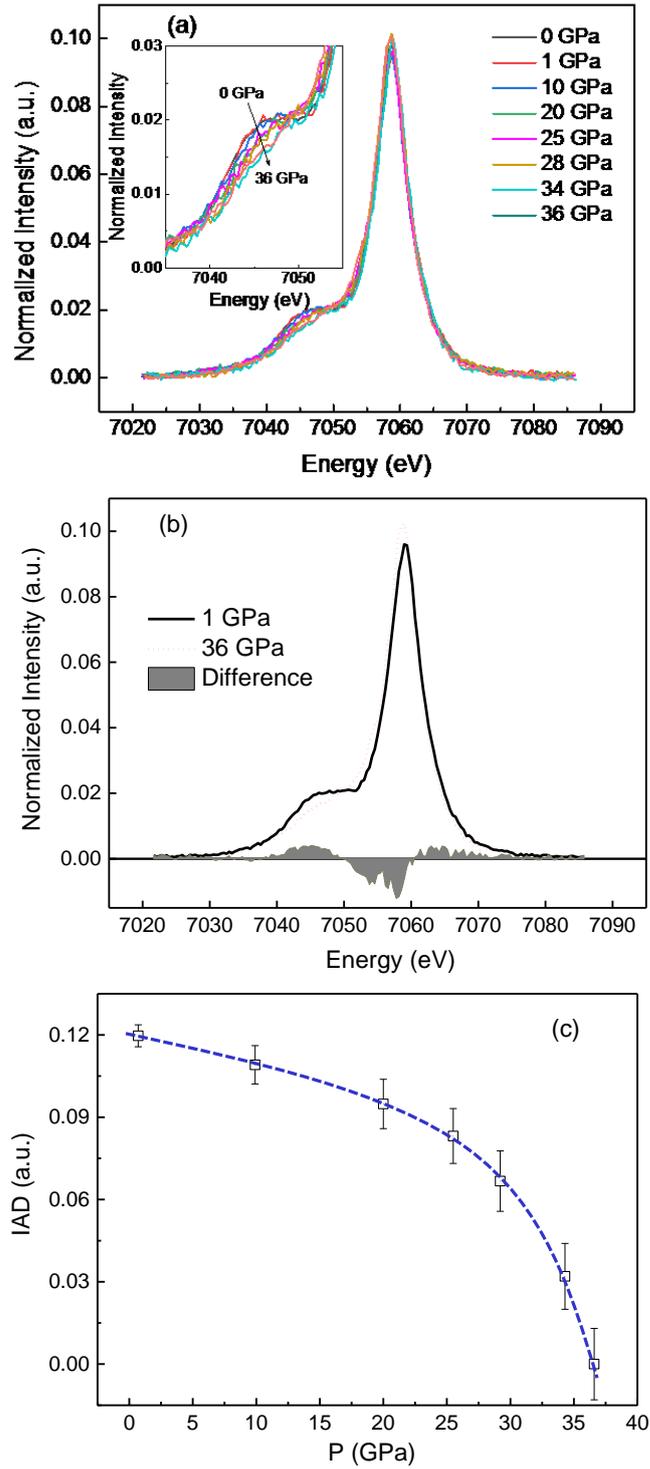

Fig. 5 (a) The X-ray emission spectroscopies of $Ba_9Fe_3Se_{15}$ measured under different high pressures at room temperature. (b) The typical normalized XES of Fe Kβ spectral line taken at 1 GPa and 36 GPa, and their difference of the spectral intensity. (c) The pressure dependence of IAD value.

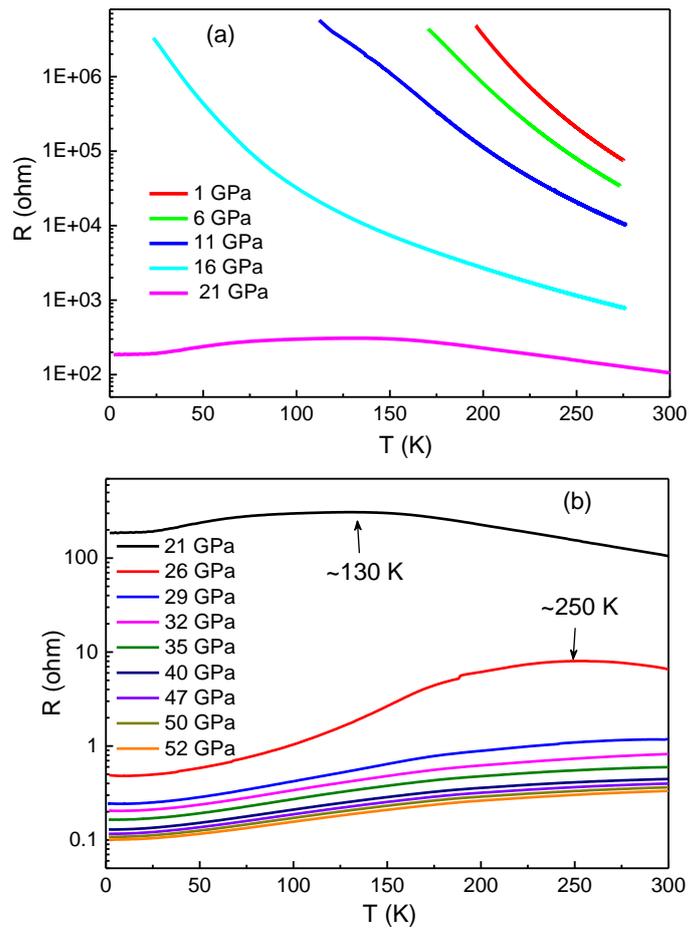

Fig. 6 The resistance measured under different pressures with the highest experimental pressure of 51 GPa. (a) The pressure range of 1-21 GPa. (b) The pressure range of 21-52 GPa.